\documentclass{aastex61}
\pdfoutput=1

\usepackage{graphicx}
\usepackage{amssymb}
\usepackage{courier}
\usepackage{epstopdf}
\usepackage{color}

\received{July 1, 2016}
\revised{September 27, 2016}
\accepted{\today}
\submitjournal{PASP}

\shorttitle{ Monitoring Telluric Absorption with CAMAL}
\shortauthors{Baker et al.}

\begin{document}

\title{Monitoring Telluric Absorption with CAMAL}

\correspondingauthor{Ashley D. Baker}
\email{ashbaker@sas.upenn.edu}

\author[0000-0002-0786-7307]{Ashley D. Baker}
\affiliation{University of Pennsylvania 
Department of Physics and Astronomy,  
209 S 33rd St, 
Philadelphia, PA 19104, USA}
\author{Cullen H. Blake}
\affiliation{University of Pennsylvania 
Department of Physics and Astronomy,  
209 S 33rd St, 
Philadelphia, PA 19104, USA}
\author{David H. Sliski}
\affiliation{University of Pennsylvania 
Department of Physics and Astronomy,  
209 S 33rd St, 
Philadelphia, PA 19104, USA}

\begin{abstract}

Ground-based astronomical observations may be limited by telluric water vapor absorption, which is highly variable in time and significantly complicates both spectroscopy and photometry in the near-infrared (NIR). To achieve the sensitivity required to detect Earth-sized exoplanets in the NIR, simultaneous monitoring of precipitable water vapor (PWV) becomes necessary to mitigate the impact of variable telluric lines on radial velocity measurements and transit light curves. To address this issue, we present the Camera for the Automatic Monitoring of Atmospheric Lines (CAMAL), a stand-alone, inexpensive six-inch aperture telescope dedicated to measuring PWV at the Fred Lawrence Whipple Observatory on Mount Hopkins. CAMAL utilizes three narrowband NIR filters to trace the amount of atmospheric water vapor affecting simultaneous observations with the MINiature Exoplanet Radial Velocity Array (MINERVA) and MINERVA-Red telescopes. Here we present the current design of CAMAL, discuss our data analysis methods, and show results from 11 nights of PWV measurements taken with CAMAL. For seven nights of data, we have independent PWV measurements extracted from high-resolution stellar spectra taken with the Tillinghast Reflector Echelle Spectrometer (TRES) also located on Mount Hopkins. We use the TRES spectra to calibrate the CAMAL absolute PWV scale. Comparisons between CAMAL and TRES PWV estimates show excellent agreement, matching to within 1 mm over a 10 mm range in PWV. Analysis of CAMAL's photometric precision propagates to PWV measurements precise to better than 0.5 mm in dry (PWV \textless 4 mm) conditions. We also find that CAMAL-derived PWVs are highly correlated with those from a GPS-based water vapor monitor located approximately 90 km away at Kitt Peak National Observatory, with a root mean square PWV difference of 0.8 mm. 

\end{abstract}

\keywords{techniques: photometric -- instrumentation: photometers -- planets and satellites: atmospheres}





\section{Introduction}

Earth's atmosphere poses a persistent challenge to astronomers making ground-based observations at near infrared (NIR) wavelengths, where telluric water vapor absorption lines are numerous. Furthermore, the strengths of these absorption lines are variable in time and depend on the amount of precipitable water vapor (PWV) present in the atmospheric column above the observatory. 

Mitigating the effects of water vapor absorption in astronomical data typically involves differential measurement techniques. While differential photometry removes first order effects of atmospheric variability, second order effects can be large depending on the spectral energy distributions of the target and reference stars. Because red stars incur more absorption than blue ones, a mismatch in the temperatures of the target star and the reference stars can still leave second-order photometric errors of over 1\% \citep{Ivezic07,Li16,Stubbs07,blake08,Blake11}. Ground-based follow-up studies of exoplanet detections around M-dwarf host stars and future surveys aiming for sub 1\% photometric precision in the NIR are among those observations most affected. For example, the MEarth survey, which is performing a ground-based search for exoplanets around M-dwarfs, rarely have comparison stars in their field-of-view (FOV) as red as their targets. As a result, the common-mode of their light curves undergo fluctuations from anywhere between 1 and 10 mmag between nights that correlate with ground-level humidity measurements \citep{Berta12}. 

Spectroscopic detection of exoplanets via the radial velocity (RV) method is also affected by telluric lines. While the large patches of telluric absorption features imprinted on the data can be masked out, \cite{Cunha14} found that micro-telluric lines can still limit RV measurements at the level of 1 m/s depending on the RV of the target star and the PWV during the observation. These microtelluric lines therefore cannot be ignored by future exoplanet RV surveys aiming to reach 10-20 cm/s precision with instruments such as ESPRESSO \citep{espresso} and NEID \citep{neid,neidhalverson}.

Radiative transfer models of Earth's atmosphere scaled according to a single optical depth parameter do an excellent job of describing the wavelength-dependent transmission of the atmosphere (e.g. \citet{Blake11}). However, a real-time measurement of the optical depth, $\tau$, is still needed to produce an appropriate atmospheric model corresponding to a specific time. While this optical depth parameter can be extracted from the high resolution science spectral data itself \citep{Artigau14, Brogi14, Osip07}, telluric lines can be blended with stellar lines. Additionally, spectroscopic data solely for the purpose of atmospheric characterization is difficult to obtain, especially if it needs to be coincident with photometric observations. 

Monitoring the PWV in real-time using a dedicated instrument would enable the generation of accurate atmospheric transmission models and alleviate the need to sacrifice valuable observing time for calibration observations. While GPS monitoring systems are commonly used for measuring PWV, these instruments require maintenance and are less accurate in dry conditions \citep{buehler12,hagemann2003}, which are typical of observing sites. An alternative to the GPS monitoring system is a multiband photometer \citep{Stubbs07}. Recently, a team of scientists from the Dark Energy Survey (DES) built a multiband photometer using narrowband filters called the aTmCam to monitor telluric absorption by constantly observing a bright star that serves to backlight the atmosphere \citep{Li13,Li14}. 

Here, we present results from a similar instrument, called the Camera for the Automatic Monitoring of Atmospheric Lines (CAMAL) that we built for use by the MINiature Exoplanet Radial Velocity Array (MINERVA) \citep{minerva15}, MINERVA-Red \citep{minervared}, and MEarth \citep{mearth} surveys located at the Fred Lawrence Whipple Observatory (FLWO) on Mount Hopkins in Arizona. CAMAL focuses on measuring PWV, which is the most important variable atmospheric absorption component for the exoplanet surveys CAMAL is aiding. In \S \ref{sec:require} we discuss the system requirements for CAMAL. We present the design of CAMAL in \S \ref{sec:Instrum} then discuss the observations taken in \S \ref{sec:data}. The methods for extracting PWV from the data are described in \S \ref{sec:extract} and in \S \ref{sec:uncertainty} we address the absolute and relative uncertainties affecting CAMAL PWV measurements. In \S \ref{sec:results} we present the results compared to nearby GPS monitors and PWVs extracted from the Tillinghast Reflector Echelle Spectrometer (TRES), which is also located on Mount Hopkins. In \S \ref{sec:conc} we summarize the performance of CAMAL and discuss future improvements to the system.

\section{System Requirements}\label{sec:require}

\begin{figure*}
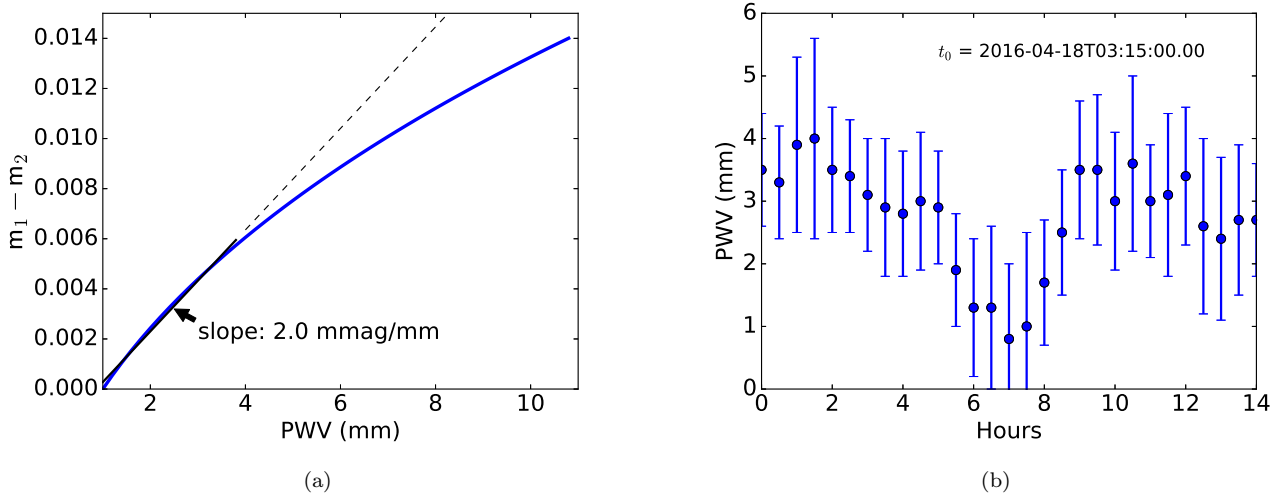

\gridline{\fig{figs/figure1a.eps}{0.48\textwidth}{(a)}
          \fig{figs/figure1b.eps}{0.48\textwidth}{(b)}
          }
\caption{Motivation for monitoring PWV in the case of the TRAPPIST-1 system. On the left we show magnitude differences as a function of PWV for a 2600 K star like TRAPPIST 1a and a 4600 K reference star. For drier conditions (PWV \textless 5 mm) the bias in differential photometry for this hypothetical observation would be 2 mmag/mm, as indicated by the slope of the linear fit. On the right we show an example time sequence of PWVs changing at Kitt Peak on April 18th, 2016.  \label{fig:TRAPPIST}}
\end{figure*}

To determine how accurate our PWV monitoring instrument must be in order to aid an exoplanet survey designed to detect transits of millimagnitude (hereafter mmag) depth, we must first determine the magnitude of the error induced by nightly PWV variations after differential photometry is performed \citep{blake08}. To model this, we assume stellar and planet parameters matching the TRAPPIST-1 system \citep{trappist} and assume a reference star of 4600 K, which is the average temperature of stars in the Tycho catalog \citep{tycho}. We use PHOENIX stellar models \citep{phoenix} for each star's spectral energy distribution and calculate these stars' fluxes for a range of PWV conditions by multiplying each stellar spectrum by a TAPAS water vapor transmission spectrum  \citep{Bertaux14} appropriate for Mount Hopkins and integrating over the effective passband of the MEarth survey \citep{mearthpassband}. In Figure \ref{fig:TRAPPIST}a we plot the magnitude difference between TRAPPIST-1a (2600K) and the 4600K reference star as a function of PWV. For relatively dry conditions (PWV \textless 5 mm), the photometric error induced by PWV absorption is 2 mmag/mm, as indicated by the black line in Figure \ref{fig:TRAPPIST}a, which is a linear fit for PWV $<$ 5 mm. In situations where there only exists reference stars hotter than 4600 K, the second order atmospheric effects will be larger than 2 mmag per millimeter change in PWV. The TRAPPIST-1 planet transit depths range from 3 mmag to 8 mmag, meaning millimeter changes in PWV could produce spurious transit-like signals or potentially obscure a true transit signal. 

How PWV actually evolves over the transit duration will ultimately determine whether the transit event will be detectable or not. For an hour long transit, which is the typical transit duration for the TRAPPIST planets, 1-2 millimeter changes in PWV can occur even during drier nights (PWV \textless 5 mm). We show an example of this in Figure \ref{fig:TRAPPIST}b using Kitt Peak PWV values from the SuomiNet network \citep{Suominet}. In this example the PWV above Kitt Peak dropped from 3 mm to 1 mm then increased back to 3.5 mm over the course of approximately four hours. An event like this could create a 4 mmag transit-like photometric dip for a star like TRAPPIST-1a, showing the importance of monitoring PWV to avoid confusing transit events for PWV variability. Though trends in PWV such as the one shown in Figure \ref{fig:TRAPPIST}b will not occur every night, transit events are intrinsically infrequent and thus monitoring PWV will help minimize the chance of either false or non-detections due to changes in PWV.

Based on these simulations, we conclude that correcting for PWV variations is crucial for precise NIR differential photometry of cool stars. Specifically, contemporaneous PWV measurements with a relative precision of $\pm 0.5$mm should provide a means to directly correct photometric measurements for second order extinction effects and achieve mmag precision. Other methods for correcting PWV variations such as using the common-mode trend of a multitude of photometry \citep{Berta12} may also reduce the impact of these second order extinction effects. While a discussion of the optimal method for making this correction is beyond the scope of this paper, we have designed CAMAL to achieve PWV measurement precision of $\pm 0.5$mm.

\vspace{5mm}
 
\section{CAMAL Instrument Design}\label{sec:Instrum}

\begin{figure}
  \centering
    \includegraphics[width=0.95\textwidth]{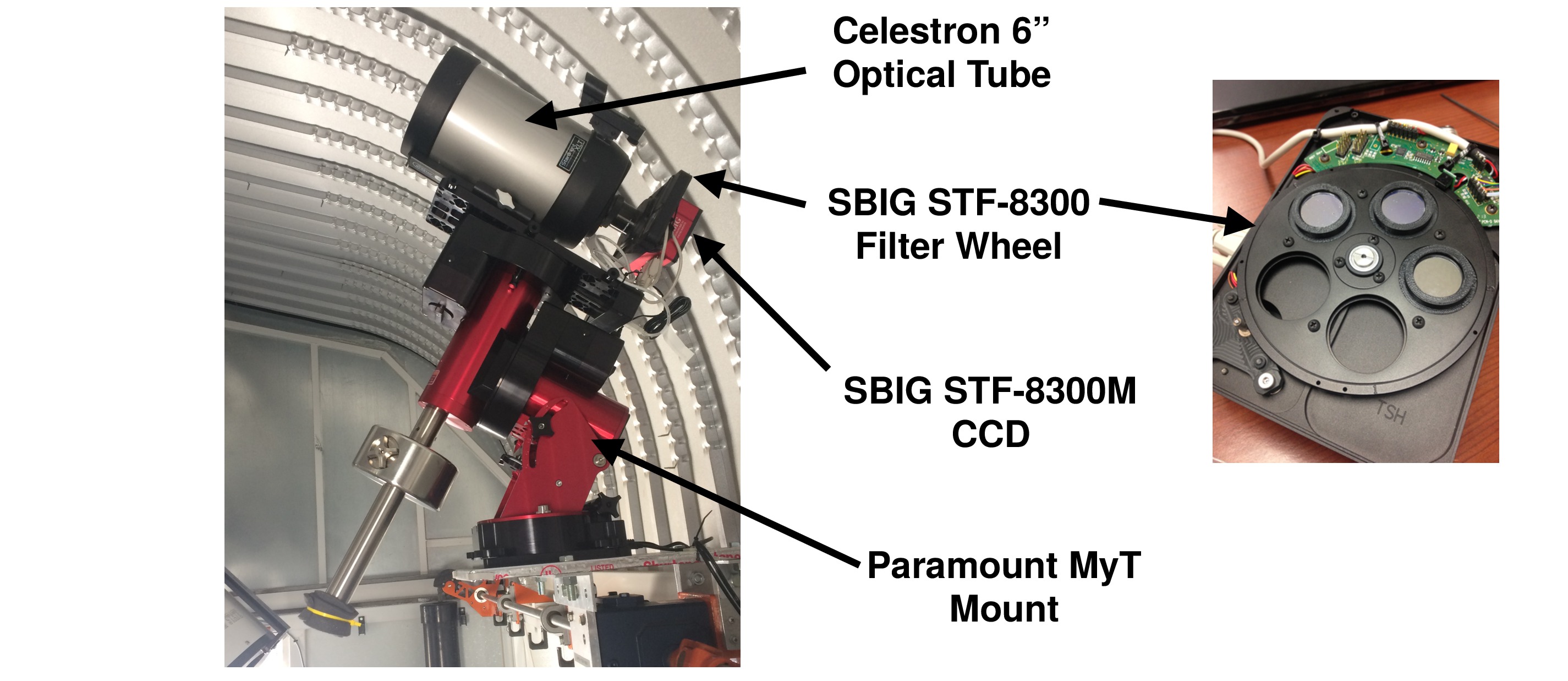}
    \caption{CAMAL is shown inside a dome housing two MINERVA telescopes. The key CAMAL components are labeled. On the right, the three CAMAL filters are shown in their custom filter holders in the SBIG filter wheel.}
    \label{fig:camal}
\end{figure}

CAMAL utilizes a Celestron six-inch aperture Schmidt-Cassegrain optical tube mounted on a robotic Software Bisque German Equatorial MyT mount. For imaging we use an SBIG STF-8300 monochrome CCD and a filter wheel that contains three narrowband filters, one of which overlaps a dense set of atmospheric water absorption lines. The other two filters are chosen to overlap regions of very low atmospheric absorption. We use the Software Bisque telescope driver along with Python scripts to automate the instrument's pointing and tracking through Software Bisque's TheSkyX control software. Figure \ref{fig:camal} shows an annotated picture of the instrument in its permanent home on Mount Hopkins in one of the MINERVA domes, in addition to an internal view of the filter wheel with the three CAMAL filters in place.

\begin{figure}
  \begin{center}
    \includegraphics[width=0.5\textwidth]{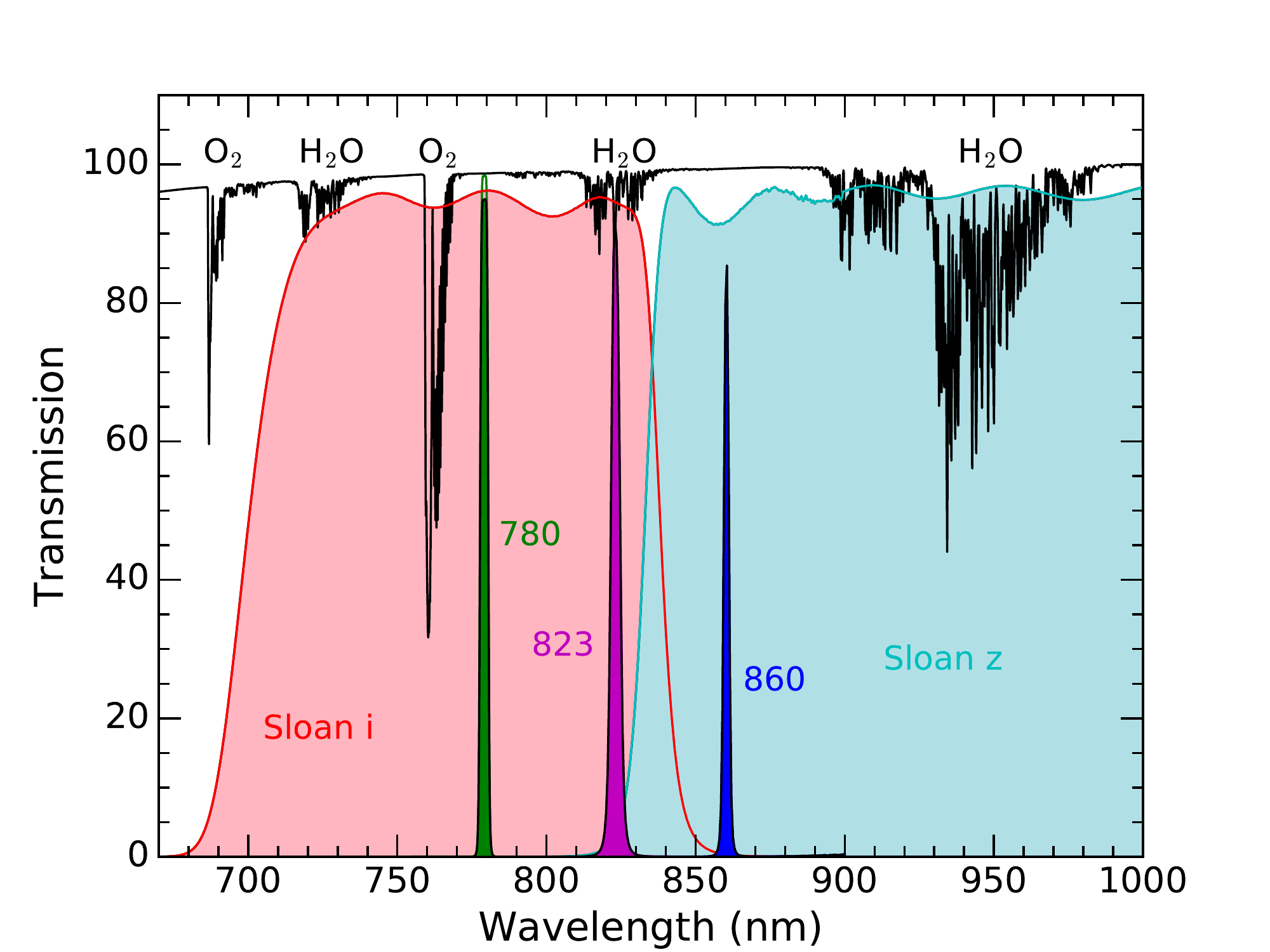}
  \end{center}
    \caption{SDSS and CAMAL filter profiles with Earth's atmospheric transmission spectrum. SDSS i and z filter profiles overlap strong water absorption features at 820 nm and 940 nm.}
    \label{fig:filters}
\end{figure}

We chose filters 25 mm in diameter with central wavelengths of 780 nm, 823 nm, and 860 nm. The 780 nm filter was an in-stock laser cleanup filter purchased from Semrock\footnote{https://www.semrock.com}, while the other two filters were production overrun filters from Omega Optical\footnote{http://www.omegafilters.com}. We list the filter properties in Table \ref{tab:filters} and plot the filter profiles over a model atmospheric transmission spectrum in Figure \ref{fig:filters}. 

We selected these filters to have central wavelengths within 100 nm of each other with the central filter overlapping water vapor absorption lines while the other two filters avoid overlapping any significant atmospheric features. Given a stellar source that serves as a backlight to the atmosphere, changes in the central 823 nm filter flux will correspond to changes in PWV, while the other filters pick up any achromatic changes in flux due to cloud coverage, changes in airmass, or intrinsic stellar variation. The three filters also were chosen to have similar physical sizes and similar narrow bandwidths on the order of a few nanometers. These requirements ensure that the exposure times for imaging our target stars are similar in each of the three filters. The narrow FWHM of these filters also ensures that the 823 nm filter is localized to a strong set of water absorption features. The 860 nm filter does not overlap with the 866.2 nm ionized calcium triplet line, which is often found in the spectra of F, G, and K type stars and can vary in time \citep{Calcium2000}. While CAMAL's targets will tend to be blue stars (T$_{\rm{Eff}}>9000$K) to avoid strong stellar features, in order to keep exposure times per filter below 10s, our limiting $i$-band magnitude is about $i$ \textless 2.5, making it necessary to occasionally observe cooler stars to remain above an airmass limit or to avoid the Moon, for example. 

Currently, CAMAL is set up to continuously observe a bright star through one filter at a time while cycling through the filter wheel. Originally, the instrument was designed to have a simultaneous imaging setup, where the filters were located at the front aperture of the telescope with wedge prisms to deviate the locations of the stellar images in the different filters on the focal plane. This was similar to the design envisaged by \cite{Stubbs07} and tested in \cite{Li13}. \cite{Li14} ultimately abandoned this design due to different optimal exposure times for each filter and filter irregularities. While our filters were chosen to ensure similar exposure times, we found that the two filters from Omega Optical were not optically flat, causing a very large change in the focus position between filters when the filters were placed at the telescope pupil. This focus difference and a lack of other filter options led us to modify our design to incorporate a filter wheel that would quickly cycle through each filter. While we may upgrade in the future to multiple telescopes and CCDs to enable truly simultaneous imaging in each filter, the filter wheel transitions are fast enough such that stars brighter than $i$-band magnitude of 2.5 can be observed with a complete filter cycle in less than 30s. \cite{Li14} analyzed the temporal variations of PWV over several nights and found that in three minutes, the PWV changes were mostly attributed to instrumental uncertainty. Although the \cite{Li14} results are from a different observing site, we feel it is safe to assume that the 30s time-scale for a filter cycle is short compared to the expected timescale of atmospheric PWV variations above Mt. Hopkins. 

Though PWV is not expected to change significantly during the timescale of one complete filter cycle, the filter wheel setup is susceptible to changes in atmospheric opacity due to nonuniform, patchy clouds moving and changing on minute timescales. In our test observations, only one night was considered to be photometric throughout, but overall we found non-photometric conditions (i.e. thin clouds) did not significantly impact our ability to extract PWV estimates from the CAMAL fluxes. The impact of clouds on CAMAL's PWV measurements is addressed in detail in \S \ref{sec:clouds}. 

\begin{table}
	\centering
	\begin{tabular}{cccc} 
		\hline
		Centers (nm) & FWHM (nm) & Peak Transmission (\%) & Vendors\\
		\hline
		779.5 & 2.8 & 88.2 & Semrock\footnote{Due to a blue leak in the 780 nm filter, it was combined with a high-pass filter from Omega Optical. The peak transmission for the 780 nm filter therefore includes the 96.5\% decrease in maximum throughput due to the addition of the high-pass filter.}\\
		823.1 & 3.2 & 91.4 & Omega\\
		860.3 & 1.9 & 85.4 & Omega\\
		\hline
	\end{tabular}
	\caption{Center wavelength, full width at half maximum, peak transmission, and vendor for each of the three CAMAL filters.}
	\label{tab:filters}
\end{table}

\section{Observations \& Data Reduction}\label{sec:data}
\subsection{CAMAL}
During a commissioning phase, we collected data with CAMAL over four nights in the summer of 2016 and seven nights in the late fall of 2016. In the summer the main target star was Polaris while in the fall the target list included Vega, Elnath, and Regulus. CAMAL continuously imaged each of its targets according to a preset schedule, cycling through the filters in the order of the 780 nm filter first, then the 823 nm filter, and ending with the 860 nm filter. Each night began by taking calibration images, including biases and twilight flats. Biases from each night are median combined to produce a nightly master bias frame. A stock set of three bias-subtracted, median combined, then normalized flat-fields for each filter are created. Flat-field corrections can become important because our small filters cause significant vignetting. In practice, the flat-field corrections typically have a negligible effect because the target star remains on the same few pixels in the center of the field of view (FOV) throughout the night. Science frames are bias-subtracted then flat-field corrected, then fluxes for the target star are extracted and the sky background subtracted using the \texttt{photutils} Python package \citep{photutils}. We refer to the flux values measured through each filter as $ f_{780}$, $f_{823}$, and $f_{860}$.

\begin{figure}
    \plottwo{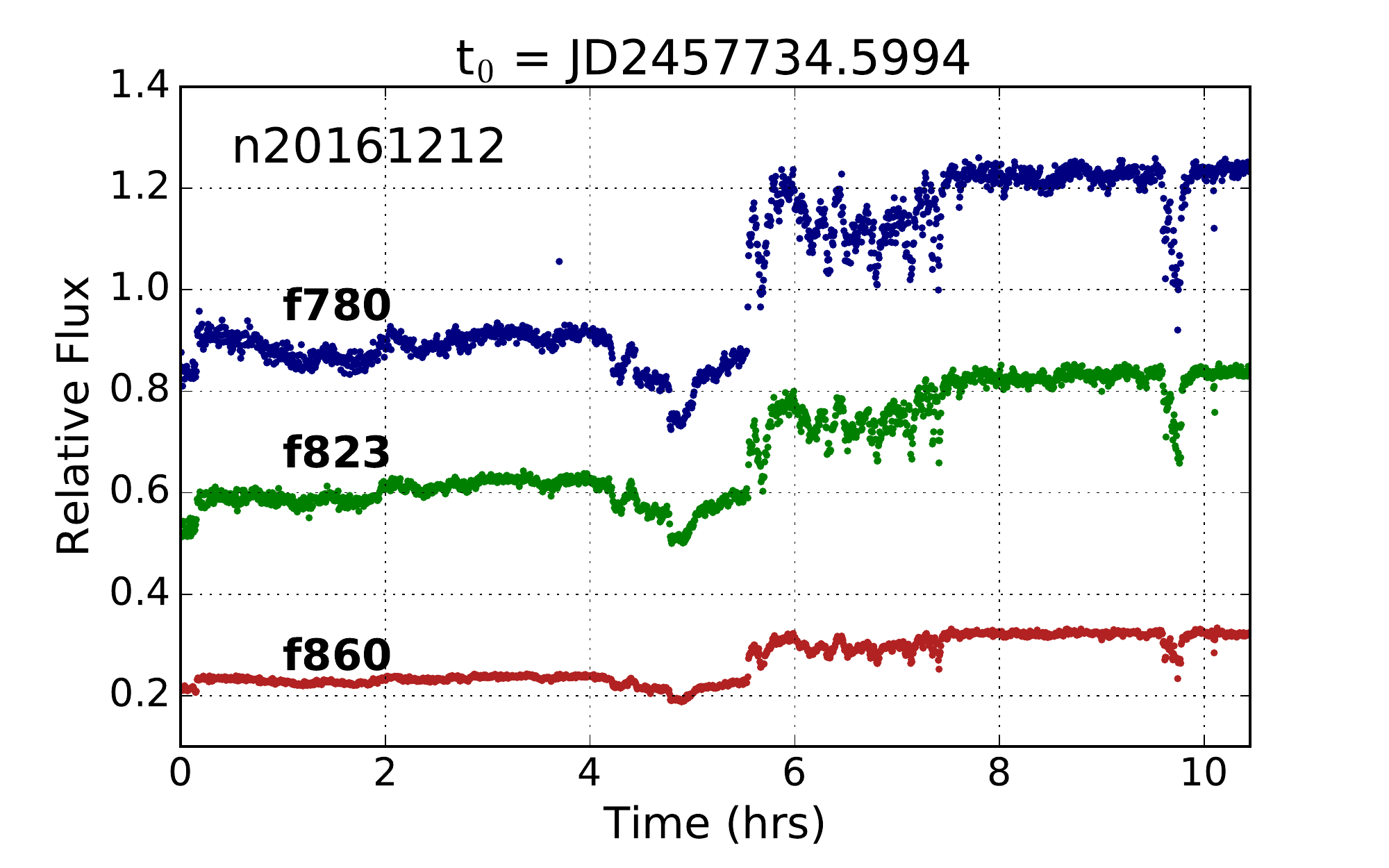}{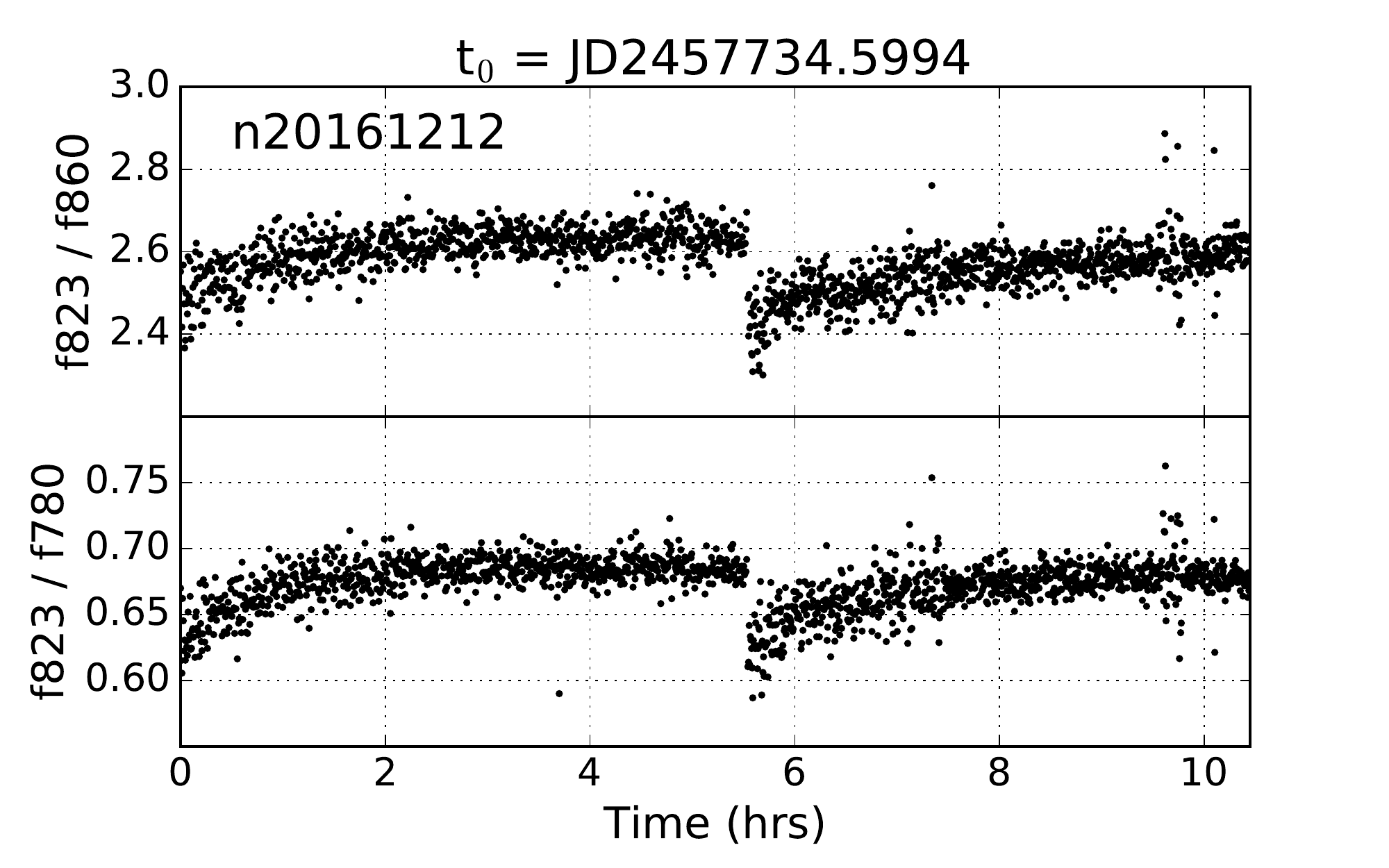}
    \caption{Example CAMAL flux time-series for the night of December 12th, 2016 (left) and CAMAL colors made from these fluxes (right). Targets switch from Elnath to Regulus at approximately 5.5 hours after the start of the run. The constant variation in the fluxes indicate that this was a very cloudy night. \label{fig:fluxexample}}
\end{figure}

We show example data from a CAMAL observing sequence in Figure \ref{fig:fluxexample}. On the left we show the relative fluxes for targets Elnath and Regulus observed on December 12th, 2016. The flux values have all been divided by a scaling factor that places $f_{780}$ around unity. The corresponding colors $c_1 = f_{823}/f_{780}$ and $c_2 = f_{823}/f_{860}$ are shown on the right in Figure \ref{fig:fluxexample}. The change in targets (and visible shift in fluxes) occurs after 5.5 hours past the start time. The increasing trend in the colors for each target star is due to the fact that the targets on the night shown here were each tracked from higher airmass (above 2) to airmass 1. Though the fluxes change by 10-20$\%$ because of clouds, this high level variability divides out when the flux ratios are taken to make the CAMAL colors.

\begin{figure}
  \centering
  \includegraphics[width=0.45\linewidth]{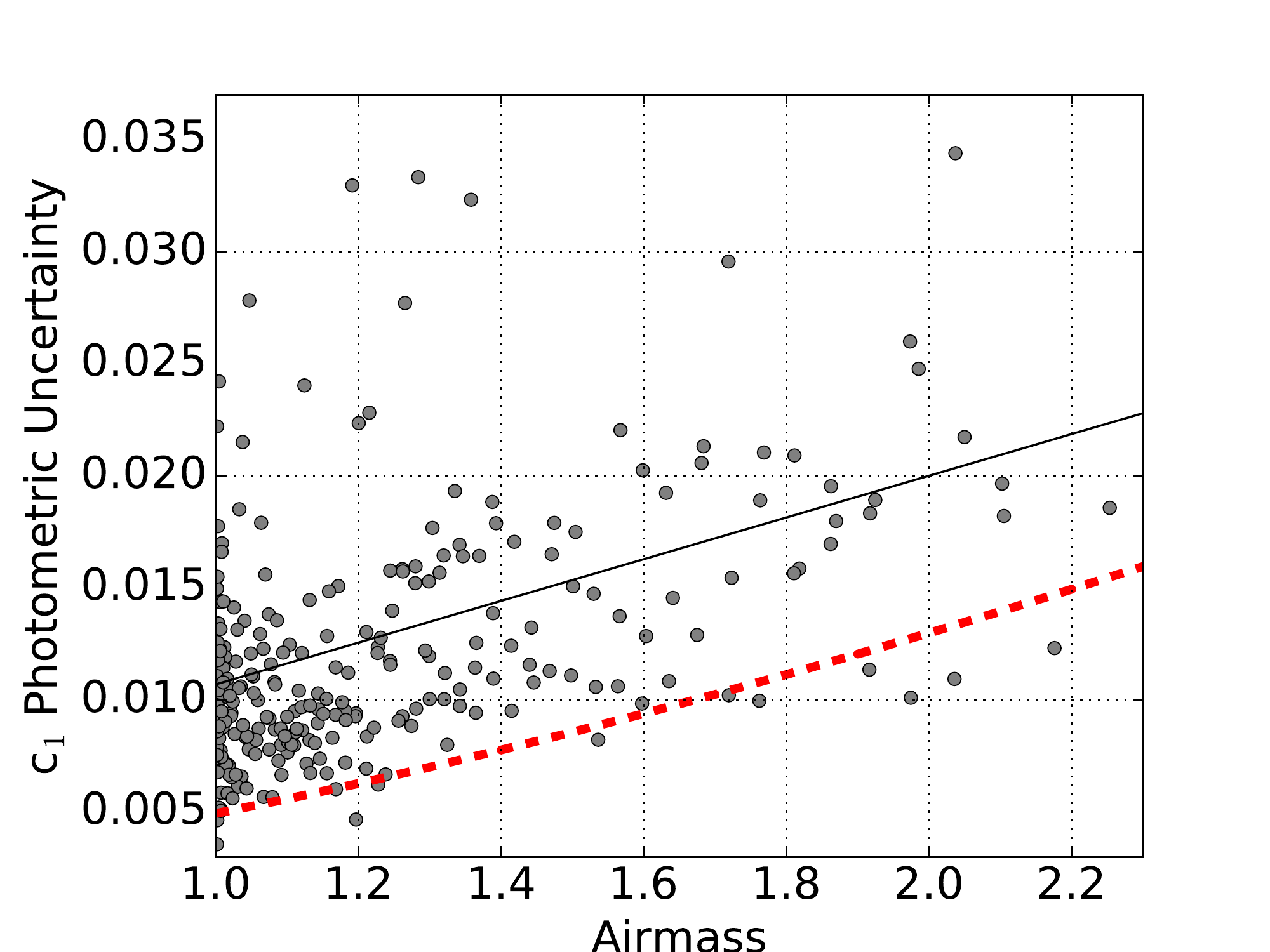}
  \caption{The RMS of five minute intervals of Elnath $c_1$ measurements plotted as a function of airmass. Five nights of data are shown here. The expected error is plotted as the red dashed line and takes into account photon noise, scintillation noise, and CCD read noise.}
  \label{fig:photunc}
\end{figure}

To assess the CAMAL photometry, we evaluate the expected photon error and compare this to the measured photometric error for the CAMAL color, $c_1$. To do this we use five nights of observations of Elnath and compute the standard deviation of $c_1$ over five minute intervals (10-11 filter cycles), which we plot as a function of airmass in Figure \ref{fig:photunc} in addition to the expected photometric uncertainty, which is shown as the red dashed line. The expected uncertainty is calculated by adding in quadrature the expected photon noise, scintillation noise, and read noise using the nominal $f_{780}$ and $f_{823}$ Elnath photon counts and then propagating this to the uncertainty in $c_1$. The estimate for scintillation noise is based on Young's approximation \citep{young67}. On average, the measured photometric error is twice the theoretical expectation based on scintillation, photon, and read noise. This underestimation may be  due to an underestimation of the scintillation noise, which has been addressed by \citet{scint1} who found that Young's approximation underestimates the median scintillation noise by a factor of 1.5. 

\subsection{TRES}\label{sec:tresdata}

Additional observations from the Tillinghast Reflector Echelle Spectrograph (TRES) on the 1.5-meter Tillinghast telescope at FLWO on Mount Hopkins are used to validate the CAMAL PWV measurements and calibrate the absolute CAMAL PWV scale. TRES is located approximately 100 meters from CAMAL and produces high-resolution (R $\sim$ 30,000) stellar spectra over wavelength ranges covering the 823 nm water absorption features. The TRES spectra were reduced and blaze corrected according to the standard pipeline for TRES spectra \citep{tres}. We use these spectra to extract PWV values directly from water line depths and compare to nearly contemporaneous CAMAL PWV measurements. The TRES instrument is on-sky in cycles of approximately two weeks as it alternates with another instrument installed on the Tillinghast telescope. Due to this, there is not always overlap between TRES and CAMAL observations. 

The TRES spectra that were selected were chosen to be the bluest, brightest star observed each night in order to reduce the influence of stellar lines on the fitting of the telluric water lines and additionally to ensure the spectra have high signal to noise ratios. Vega was a common target, but because Vega was typically observed during twilight several hours before CAMAL started observing, we also utilized TRES spectra of other targets that were observed closer in time to the CAMAL observations. In total, we utilize eight TRES spectra that overlap with seven CAMAL observing nights. Not all TRES observations are exactly contemporaneous with CAMAL observations because TRES begins observations of the brightest stars well before astronomical twilight. The airmasses of the TRES observations are also different than the CAMAL observations. We discuss this more when comparing CAMAL and TRES PWV measurements in \S \ref{sec:results}. 

\begin{figure}
  \begin{center}
    \includegraphics[width=0.6\textwidth]{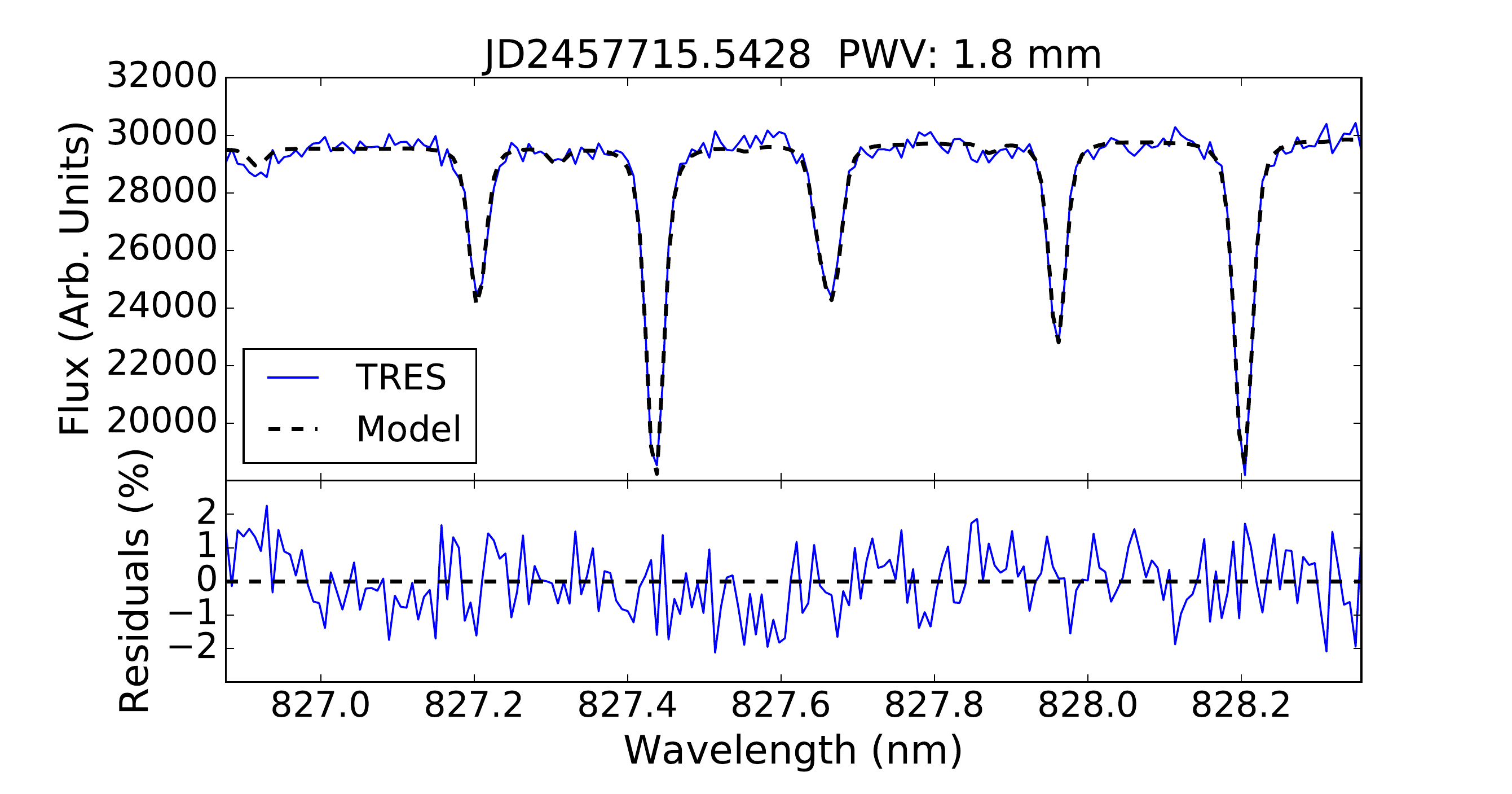}
  \end{center}
    \caption{Example fit of a TRES spectrum of Vega on November 23rd, 2016. The top panel shows the spectral data in blue with the best fit plotted as the black dashed line. The bottom panel shows the residuals of the fit. }
    \label{fig:tresfit}
\end{figure}

To extract PWV from the TRES data, we fit a 15\AA $\:$portion of the spectrum around 828 nm. We use a model atmospheric water vapor transmission spectrum appropriate for Mt. Hopkins from TAPAS \citep{Bertaux14} to fit the data. We scale the water vapor transmission spectrum to a particular PWV value following Equation \ref{eq:PWV}. We then convolve the model spectrum with a Gaussian of width $\sigma$ to model the line spread function of the spectrograph, then multiply by a linear function to account for a slope in the continuum. Two parameters are used to describe the wavelength solution of the data, then the model spectrum and data are compared on the same wavelength grid. We use the \texttt{MC3} Python package to find the best fit parameters and parameter uncertainties. Since we do not account for airmass in the fit, we must divide the final best fit PWV value by the airmass of the observation to scale PWV to its value at an airmass of 1. This allows for a direct comparison to CAMAL PWVs, which also determines PWV scaled to an airmass of 1 (see Equation \ref{eq:flux}). An example fit of a TRES spectrum is shown in Figure \ref{fig:tresfit}. For stars that contain more spectral absorption features, we fit multiple 15\AA $\:$ regions and average the results for PWV. 

\section{Extracting PWV from the CAMAL Colors}\label{sec:extract}
To derive absolute PWV measurements from the CAMAL photometry, we construct a forward model that produces model CAMAL fluxes by integrating over a template stellar spectrum multiplied by the throughput of the entire optical system in each filter band. The forward model is pre-evaluated on a three dimensional grid of PWVs, airmasses, and stellar temperatures. Then, for a specific  airmass and target-star temperature, the PWV can be quickly evaluated by matching the observed CAMAL fluxes to those generated by the forward model. This quick calculation is useful when real-time PWV measurements are required. Alternatively, the time sequence of PWV over the night can be represented as a smoothly varying function and the whole night's data can be fit using an MCMC approach. We use this as a second method to extract PWVs from the data that allows us to characterize the statistical uncertainty in our PWV measurements.

\subsection{Generating Model Color Grids}\label{sec:pwvquick}

We make use of the instrument response curves for the optical system, CCD, and filters provided by the vendors to predict the overall wavelength dependent efficiency for CAMAL. For the model spectrum of the atmospheric absorption, we use the TAPAS \citep{Bertaux14} web form\footnote{http://www.pole-ether.fr/tapas/} to download one model spectrum containing only water vapor absorption features and another spectrum that only includes the effect of Rayleigh scattering. We scale the TAPAS telluric absorption model using $\textrm{v}$, the precipitable water in a column of unit area in the atmosphere, according to
\begin{equation}\label{eq:PWV}
 \textrm{W}(\textrm{v}^\prime) =  \textrm{W}(\textrm{v})^{\textrm{v}^\prime/\textrm{v}} ,
\end{equation}

\begin{figure}
  \begin{center}
    \plottwo{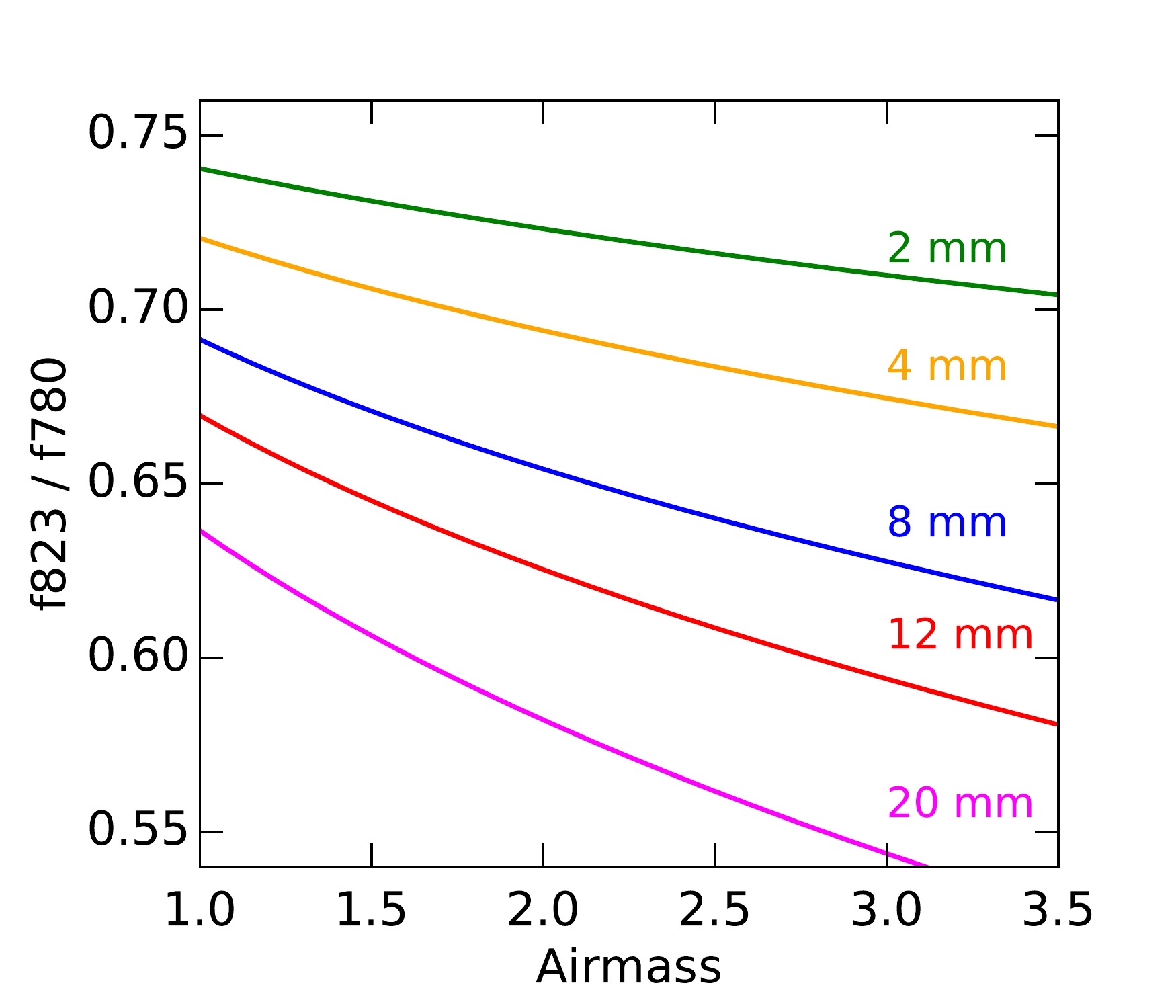}{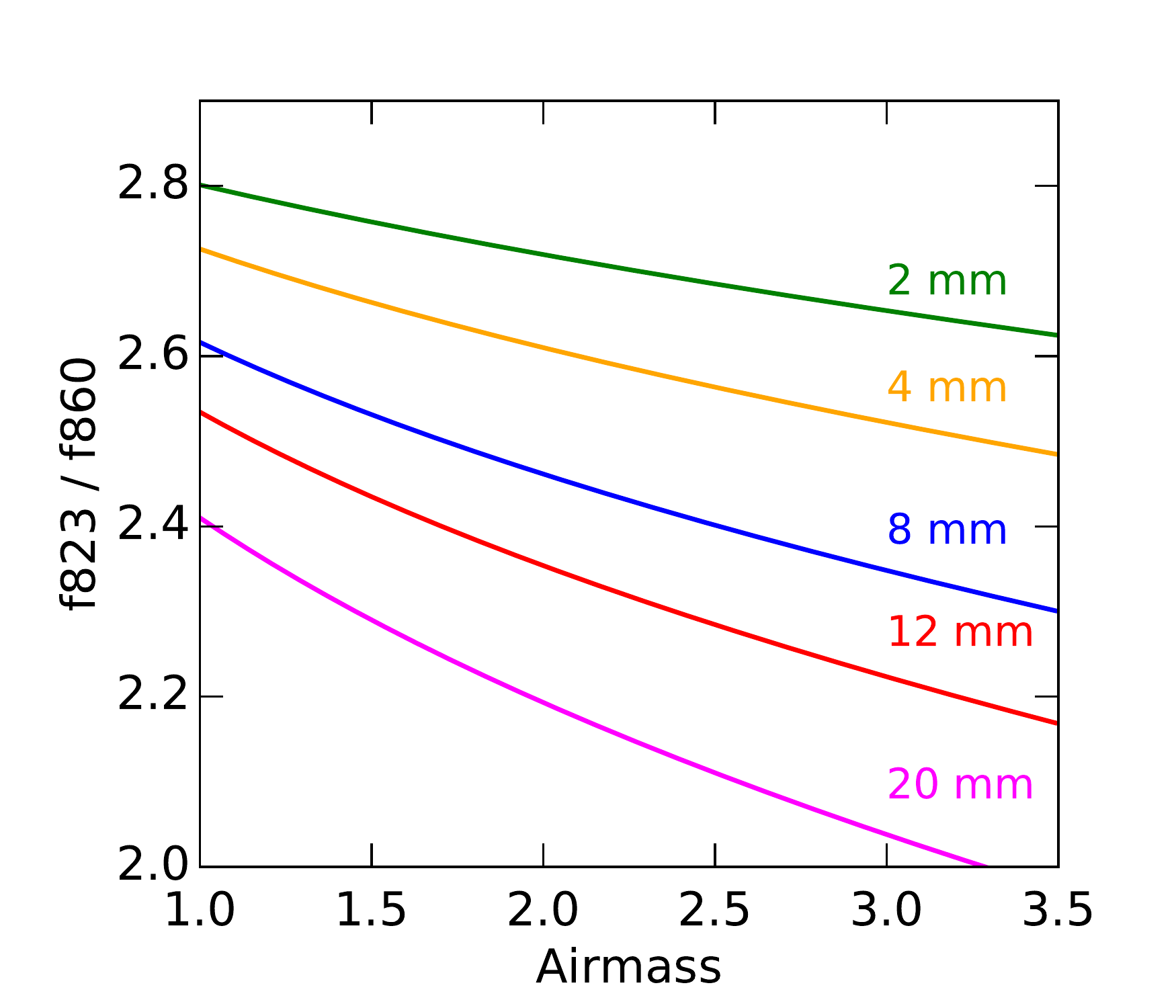}
  \end{center}
    \caption{The flux ratios for $f_{823}$/$f_{780}$ and $f_{823}$/$f_{860}$ plotted versus airmass for a 10,000 K star at different PWV values (in millimeters) as determined by the forward model. The location on this plot of the measured CAMAL color for a given observation indicates the PWV at the time of the observation.}
    \label{fig:fluxairmass}
\end{figure}

\noindent where $\textrm{W}$ is a telluric water vapor absorption spectrum generated using a known value of $\textrm{v}$. We define the parameter $\tau$, which is related to the water optical depth, as $\tau = \textrm{v}^\prime/\textrm{v}$ .

For each of the stars that are observed, we use PHOENIX \citep{phoenix} synthetic spectra of the corresponding temperature, surface gravity, and metallicity to predict the flux per unit telescope area, $f$, through each filter, $i$, which is derived using the following equation:

\begin{equation}\label{eq:flux}
f_i = \int d\lambda \: \textrm{W}^{\tau \mathit{X}}(\lambda) R^\mathit{X}(\lambda)  B_i(\lambda)  S(\lambda, T)  E(\lambda) .
\end{equation}

\noindent  We use $R$ as the Rayleigh scattering function slope at a given airmass, $X$, $B$ for the filter bandpass, $S$ as the stellar spectrum, and $E$ as the combined efficiency of the telescope and CCD. We always take PWV to be at an airmass of 1 and then scale by the airmass of an observation. In this equation, the only time-dependent variables are airmass and stellar temperature, which are both known, and $\tau$, which we wish to measure. We pre-evaluate Equation \ref{eq:flux} for all possible observations by integrating over wavelength assuming a grid of airmass values from 1.0 to 3.5 in steps of 0.01, stellar temperatures and properties matching those of our targets, and PWV values from 0 mm to 25 mm in steps of 0.1 mm. We use these model flux grids to generate the model colors $m_1 =  \frac{f_{823}}{f_{780}}$ and $m_2 = \frac{f_{823}}{f_{860}}$ to compare to the observed CAMAL colors, $c_1$ and $c_2$. For each observation of the CAMAL color of a target star, the PWV during that measurement can be determined by where the observed color is located on this grid of model colors. In Figure \ref{fig:fluxairmass} we show example model colors, $m_1$ and $m_2$, both plotted versus airmass for a 10,000 K star and various values of PWV. We find that both colors are very sensitive to PWV. At an airmass of 1.3 and PWV value of 7 mm, there is a 1\% change in $m_1$ per millimeter change in PWV. 

\subsection{Modeling PWV as a Smooth Function}\label{sec:fitting}

In the grid-based approach to extracting PWV, each CAMAL color must be matched to its respective color grid, which produces two different mappings to PWV. Although these two PWV mappings agree with each other under normal conditions (i.e. no external chromatic effects) and are just limited by photon noise to within our PWV grid spacing of 0.1 mm, we would prefer a method that fits both colors at the same time to find the best fit given all of the photometry and derives a single PWV time series for the night that is smoothly varying. To do this, we model the PWV as a smoothly varying function in time under the assumption that the typical timescale for PWV variability is 30 min \citep{Li14}. Modeling PWV in this way allows us to fit the entire night of data at once using an MCMC sampler that allows us to characterize our uncertainties. 

The fit is performed by using a spline function to define PWV as a function of time. We map this PWV time sequence to the time variation of CAMAL colors using the model color grids described in \S \ref{sec:pwvquick}. The optimal PWV function parameters are determined by minimizing the difference between the model and observed colors following value:

\begin{equation}\label{eq:min}
\sum_i \Big[ \big(c_1(t_i) - m_1(\textrm{PWV}(t_i))\big)^2 + \big(c_2(t_i) - m_2(\textrm{PWV}(t_i))\big)^2\Big] \: .
\end{equation}

\noindent Here, $c_1$ and $c_2$ represent the CAMAL observed colors, and $t_i$ is the time of exposure sequence $i$. To define PWV($t_i$) we use a 3rd order B-spline, which we define using the Python function \texttt{splev} from the \texttt{scipy.interpolate} package. More information regarding the definition of the spline can be found in the references for \texttt{splev} \citep{bspline1,bspline2,bspline3}. We force the spline knot points to be stationary and equally spaced such that there is at least one knot point in each hour interval, including the end points. The hourly spaced knot points allow for variations on 30-minute time scales without overfitting the data. If $N_{in}$ is the number of internal knots, then there are $N_{in} + 4$ coefficients whose initial values are set to a starting value between 1 and 20 mm based on the model grid estimates. We use the \texttt{MC3} Python package \citep{mc3} to find best fit coefficients that describe $\textrm{PWV}(t)$. The various realizations of $\textrm{PWV}(t)$ using the parameters in the converged MCMC chains are used to determine the uncertainty in the PWV measurements while the median of them is taken to be the best fit.


\section{Assessing PWV Uncertainty}\label{sec:uncertainty}
Depending on the application, either relative or absolute PWV estimates may be necessary for correcting NIR spectroscopic or photometric measurements. We assess CAMAL's PWV accuracy in terms of how well CAMAL can measure absolute PWV values and we discuss precision in the context of CAMAL's ability to measure relative PWV values over time. In addition to this we discuss and characterize how clouds can contribute to both types of these uncertainties. 

\subsection{Accuracy of Absolute PWV Measurement}\label{sec:accuracy}
The absolute calibration of CAMAL PWV estimates based on our forward modeling of the measured stellar flux depends on the total system throughput as a function of wavelength and the accuracy of our stellar models. However, the instrument efficiency profiles provided by the manufacturers are not direct measurements of our instrument. Additionally, while the 780 nm and 823 nm bands do not fall on any broad stellar spectral features in our hot target stars, the 860 nm band does overlap a Paschen series line that can be mismatched to our model spectrum depending on the true broadening of that line. While this could be alleviated by obtaining high-resolution spectra for each of our target stars over CAMAL's wavelength region, we can instead rely on the fact that the CAMAL flux through the 780 nm and 860 nm bands should match our model since flux through these two bands is independent of PWV. To calibrate the 823 nm model throughput, which is degenerate with PWV, we use independent PWV values extracted from TRES spectra. 

Because the 860 nm band overlaps a broad stellar spectral feature, we find that each star requires its own adjustment to its efficiency at 860 nm. The CAMAL filters are very narrow and so we can take the correction to both the instrument efficiency and stellar spectrum to be a scalar, $A_{1}$, that is specific for each star. We indicate this by making it a function of temperature. $A_1(T)$ is just the ratio of the true stellar flux and telescope efficiency (i.e. mirror reflectivity, CCD QE) at 860 nm divided by what was originally assumed:

\begin{equation}
   A_1(T) = \frac{S_{\mathrm{true}}(\lambda, T)  E_{\mathrm{true}}(\lambda)}{S(\lambda, T)  E(\lambda) }  \biggr|_{\lambda=860\rm{nm}}
\end{equation}

\noindent This factor $A_1(T)$ is determined such that $f_{780}/f_{860}$ from our model agrees with our average flux measurement at airmass 1.0 for each star. We find that $A_1(T)$ ranges between 0.88 and 0.94 for the range of targets used in this work. To assess how reasonable these offsets are, we use a true spectrum of Vega from the CALSPEC Calibration Database\footnote{http://www.stsci.edu/hst/observatory/crds/calspec.html} to calculate $f_{780}/f_{860}$ and compare this to what we estimated using a PHOENIX synthetic spectrum. We find that $f_{860}/f_{780} $ increases by 4\% when we use the real spectrum of Vega compared to the PHOENIX model. Given the uncertainties in the reported efficiencies of the system components, such as the CCD, this is consistent with $A_1$ = 93.4\%, which is what was found for Vega. 

For calibrating $f_{823}$, we find no significant difference in $f_{823}/f_{780}$ when calculated with a real and synthetic spectrum for Vega. We therefore only determine one unique factor to account for uncertainty in CAMAL's assumed instrument efficiency at 823 nm:

\begin{equation}
    A_2 = \frac{E_{\mathrm{true}}(\lambda) }{ E(\lambda) } \biggr|_{\lambda = 823 \rm{nm}}
\end{equation}

\noindent To determine $A_2$, we compare CAMAL PWV estimates to the TRES measurements of PWV. Because TRES PWVs are extracted from high resolution spectra using an atmospheric model, these PWVs are unbiased, independent measurements of PWV. We compute $A_2$ by minimizing the root mean square difference between the TRES direct measurements of PWV and the corresponding PWV estimates from the CAMAL colors. We find that $A_2$ is 97.9\%, which implies that we were overestimating CAMAL's throughput by 2.1\%, which is reasonable due to the uncertainties in the telescope reflectivity and CCD QE. Using the two scalars $A_1(T)$ and $A_2$ to recalculate Equation \ref{eq:flux} and remake our model grids. We use these calibrated grids to recompute PWV for each night. A plot of the CAMAL-TRES comparison before $A_2$ is applied and after each of these two calibrations are performed is presented in Figure \ref{fig:trescamal}.

\subsection{PWV Measurement Precision}\label{sec:precision}
Assuming there are no other time varying chromatic effects in the CAMAL system other than PWV changing, the precision of CAMAL's measurements relies only on its ability to measure precise colors of its target star. We can therefore estimate its measurement precision by calculating the variation in our color measurements and then relating this to PWV measurement precision by propagating these color uncertainties through our PWV extraction method. In reality, clouds can potentially act as a time varying absorption that affects each filter slightly differently because the observations are not exactly contemporaneous. We address this in \S \ref{sec:clouds}.

\begin{figure}
\centering
\begin{minipage}{.48\textwidth}
  \centering
  \includegraphics[width=0.99\linewidth]{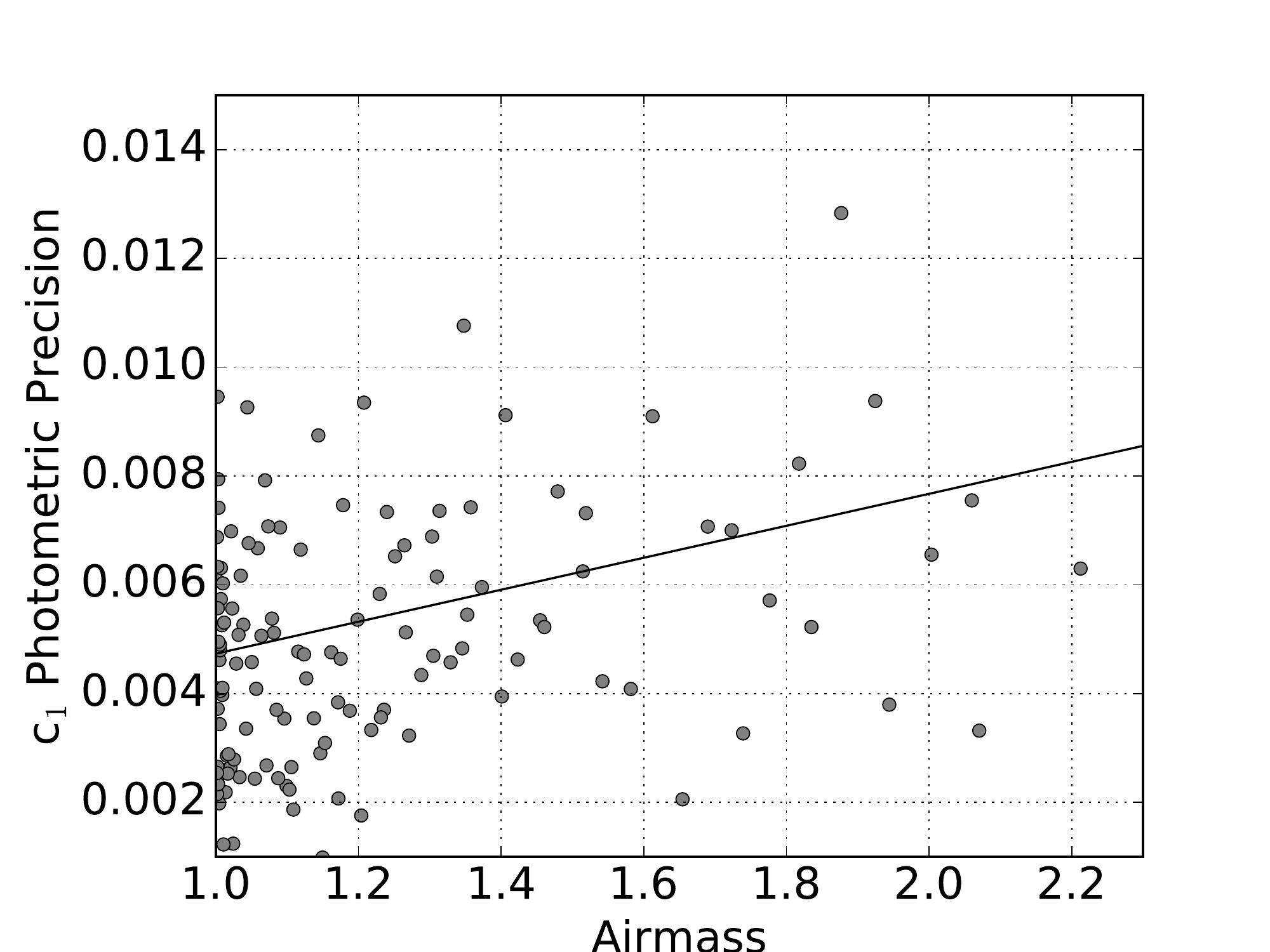}
  \caption{Empirical photometric precision of CAMAL colors as a function of airmass for five nights of observations of Elnath. }
  \label{fig:photerr}
\end{minipage}
\hfill
\begin{minipage}{.48\textwidth}
  \centering
  \includegraphics[width=0.99\linewidth]{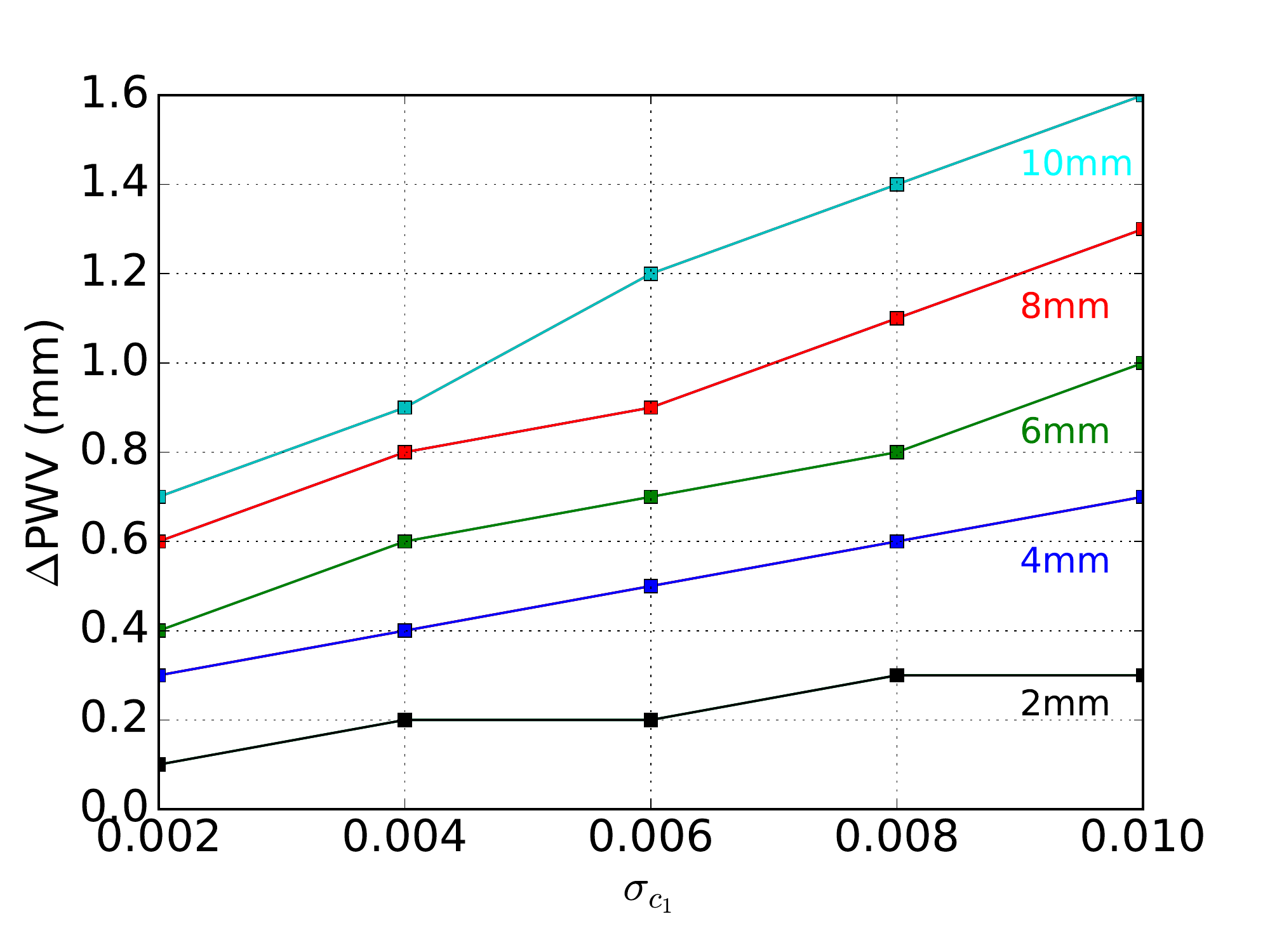}
  \caption{The error in PWV as a function of error in $c_1$ for various values of PWV assuming a 10,000K star and an airmass of 1.3. The resulting PWV error is step-like due to the 0.1 mm resolution of the model color grids. }
  \label{fig:pwverr}
  \end{minipage}
\end{figure}

Using five nights of CAMAL photometry on the target star Elnath, we estimate the photometric precision of our observations by taking the standard deviation over ten minute intervals of binned CAMAL colors $c_1$ and $c_2$. We plot the standard deviation of each of these ten minute intervals for $c_1$ as a function of airmass in Figure \ref{fig:photerr}. This differs from Figure \ref{fig:photunc} because here we have binned $c_1$ and $c_2$ on 2 minute timescales (4-5 filter cycles) to average down the photometric noise in order to see how precise we measure $c_1$ on longer timescales. For future CAMAL operations, we will limit observations to airmass less than 1.5 to reduce the effects of scintillation noise on CAMAL photometry. 

Using the empirical range in photometric precision (0.002 \textless $\sigma_{c_1}$ \textless 0.01) we found for $c_1$, we propagate this photometric uncertainty into PWV uncertainty by using our model color grids to map color to PWV. We determine $\Delta$PWV, the difference in true PWV and PWV determined after adding $\sigma_{c_1}$ to $c_1$. We plot $\Delta$PWV versus $\sigma_{c_1}$ for various PWV conditions in Figure \ref{fig:pwverr}. For this calculation we assume an airmass of 1.3 and target star temperature of 10,000K. At an airmass of 1.3, $\sigma_{c_1}$ clusters around 0.006 in Figure \ref{fig:photerr}. This propagates to a precision of better than 0.5 mm in drier conditions, when PWV \textless 4 mm. These estimates of CAMAL's PWV precision agree with the statistical uncertainties in PWV determined from the fit to the photometry using a spline functional form for PWV as described in Section 5.2.

\subsection{The Impact of Clouds}\label{sec:clouds}
Due to the filter wheel design of CAMAL, the instrument does not make perfectly simultaneous measurements in all three bands. Though filter cycles are short compared to the expected timescale of variations in the atmosphere, clouds move very quickly across the field of view and therefore can effectively introduce a chromatic effect as the cloud extinction varies over a complete CAMAL filter cycle. If the magnitude of the cloud extinction changes randomly, clouds will increase the scatter in the photometry, worsening the photometric precision. If the cloud extinction is inherently chromatic, this could also systematically bias our measurements of PWV. To understand the magnitude of this effect, we use a night of data (Dec. 6th, 2016) that was clear at the beginning of the night, was affected by thin clouds for for a period of approximately an hour, then finished with a very sharp drop in flux due to thickening clouds.

\begin{figure}
  \begin{center}
    \includegraphics[width=0.5\textwidth]{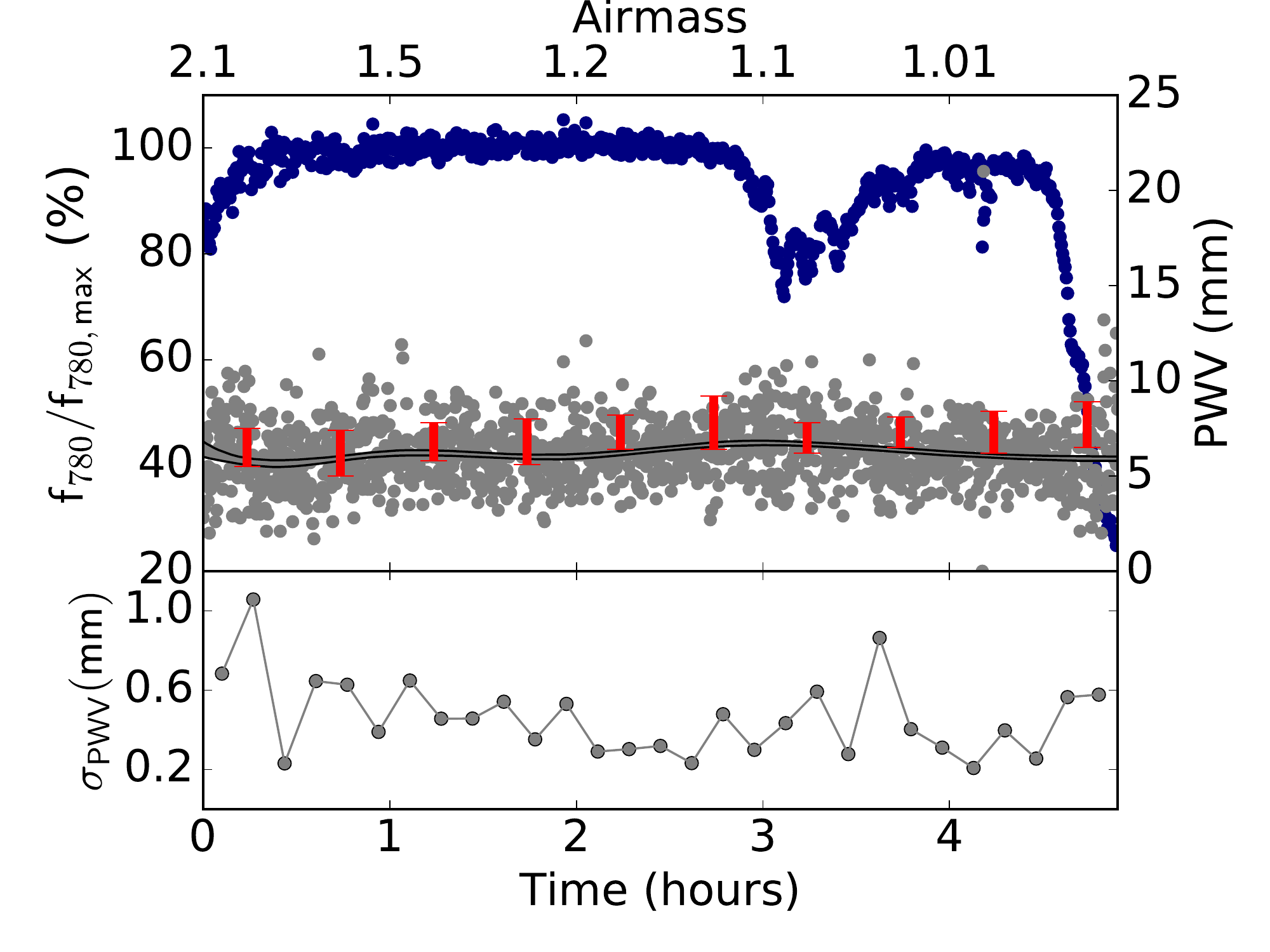}
  \end{center}
    \caption{In the top panel, the $f_{780}$ flux of Elnath on December 6th, 2016 in addition to CAMAL PWVs for the night and Kitt Peak PWVs in red. The gray points are from PWV measurements mapped from each individual CAMAL color and the best fit PWV time sequence is given as the gray line. In the bottom panel we plot PWV error estimated from the individually mapped PWVs. High airmass observations (X \textgreater 1.5, time \textless 1 hour) and regions where flux is changing due to clouds (time \textgreater 2.8 hours) show greater error.}
     \label{fig:clouderr}
\end{figure}

To understand if clouds cause a systematic bias in the estimated PWVs, we look for systematics in the extracted best fit PWV that correlate with cloud absorption. If clouds are not acting as gray absorbers, we would expect a systematic trend in PWV correlating with changes in the overall flux attenuation due to clouds. In the top panel of Figure \ref{fig:clouderr}, we plot $f_{780}$ for Elnath along with CAMAL PWVs in gray and PWVs from the nearby site at Kitt Peak National Observatory plotted in red. Where flux is decreasing due to clouds (time \textgreater 2.8 hours) we see no systematic trend in the CAMAL PWVs. While the PWV between Mt. Hopkins and Kitt Peak are not perfectly correlated (see in \S \ref{sec:GPS}) the Kitt Peak PWVs are consistent with constant for this night, implying that the overall PWV level at Mt. Hopkins was likely stable over this night as well, consistent with the CAMAL measurements shown in Figure 10. Based on this, we conclude from this that chromatic effects in clouds do not significantly bias the CAMAL photometric PWV estimates. 

To estimate how clouds affect PWV precision, we take the `real time' PWVs extracted from the model grids. We would expect increased photometric scatter due to clouds to be evident in the scatter of the grid-based PWV estimates. We show the standard deviation of ten minute intervals of the grid-based PWVs binned by 5 filter cycles (2 minutes) in Figure \ref{fig:clouderr}. The scatter in PWV decreases with time because of decreasing airmass, but increases again at times coincident with cloud absorption. While the effects of clouds are not debilitating, in future runs we will choose to move targets if cloud absorption reduces the nominal target flux by over 20\%, which will aid CAMAL in maintaining its goal of 0.5 mm precision in PWV.

\section{Results \& Discussion}\label{sec:results}

\begin{figure*}
\gridline{\fig{figs/figure11a.eps}{0.48\textwidth}{(a)}
          \fig{figs/figure11b.eps}{0.48\textwidth}{(b)}
          }
\gridline{\fig{figs/figure11c.eps}{0.48\textwidth}{(c)}
          \fig{figs/figure11d.eps}{0.48\textwidth}{(d)}
          }
\gridline{\fig{figs/figure11e.eps}{0.48\textwidth}{(e)}
          \fig{figs/figure11f.eps}{0.48\textwidth}{(f)}
          }
\caption{Data from six nights of CAMAL data taken in December, 2016. CAMAL PWVs are plotted as gray points with the best-fit function of PWV overplotted as a gray shaded region. Red and blue error bars show PWV measurements from the nearby GPS PWV monitors at Kitt Peak and Amado, respectively. Regions with high scatter in the CAMAL PWVs typically correspond to changing to a high airmass target or because of clouds passing in the FOV.\label{fig:camaldata}}
\end{figure*}

\begin{figure}
\centering
\begin{minipage}{.95\textwidth}
  \centering
  \includegraphics[width=0.57\linewidth]{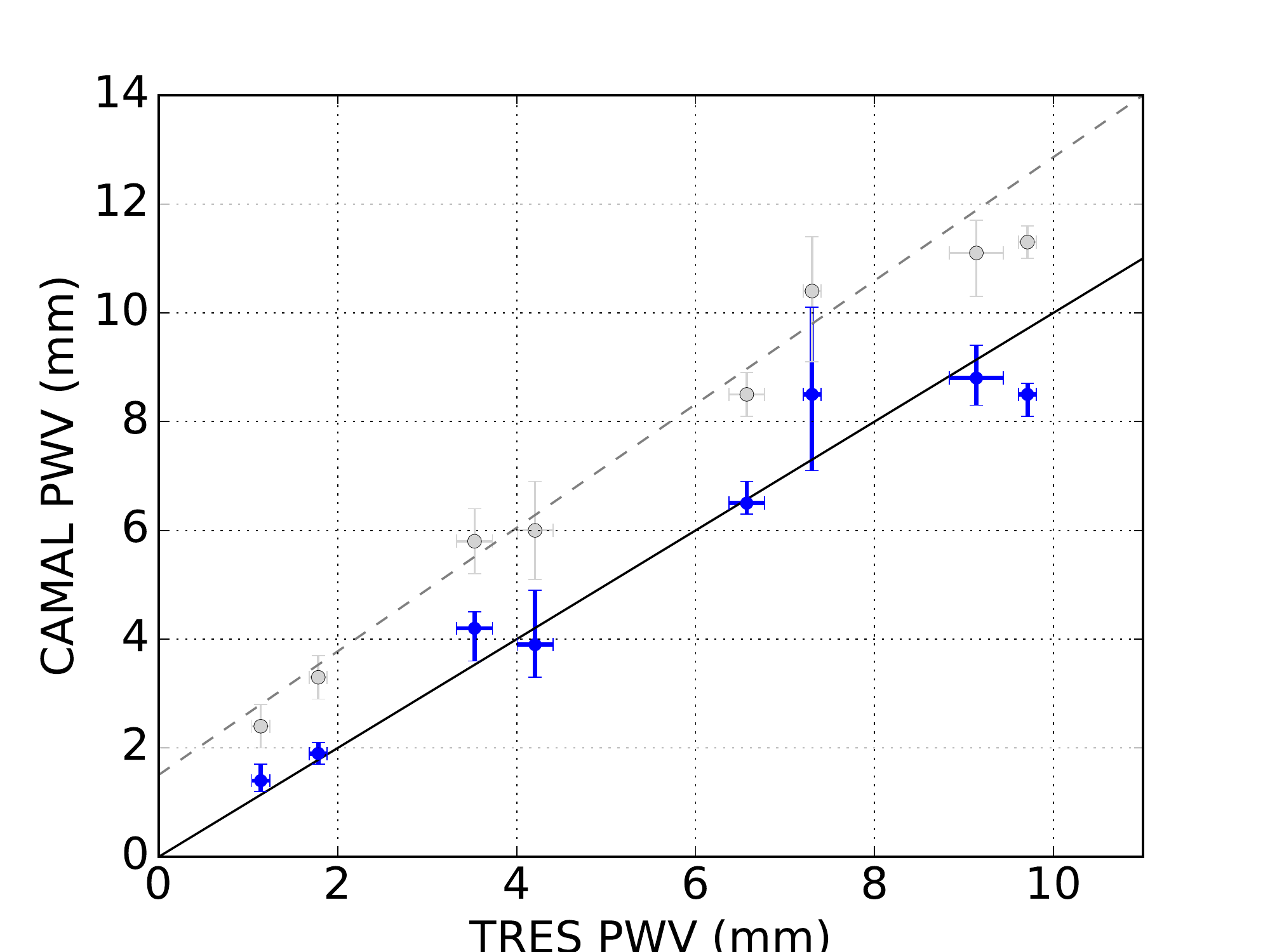}
  \caption{CAMAL PWVs plotted versus independent PWV measurements using TRES. The gray points are using CAMAL measurements before calibrating the instrument throughput. The gray dashed line is a linear fit to these data, which were used to determine the calibration factor, $A_2$. The blue points compare TRES with the CAMAL data after calibrating with $A_2$. The solid black line is a one-to-one line. The standard deviation of residuals of TRES compared to the calibrated CAMAL PWVs is 0.7mm.}
  \label{fig:trescamal}
\end{minipage}
  \hfill
\begin{minipage}{.48\textwidth}
  \centering
  \includegraphics[width=0.9\linewidth]{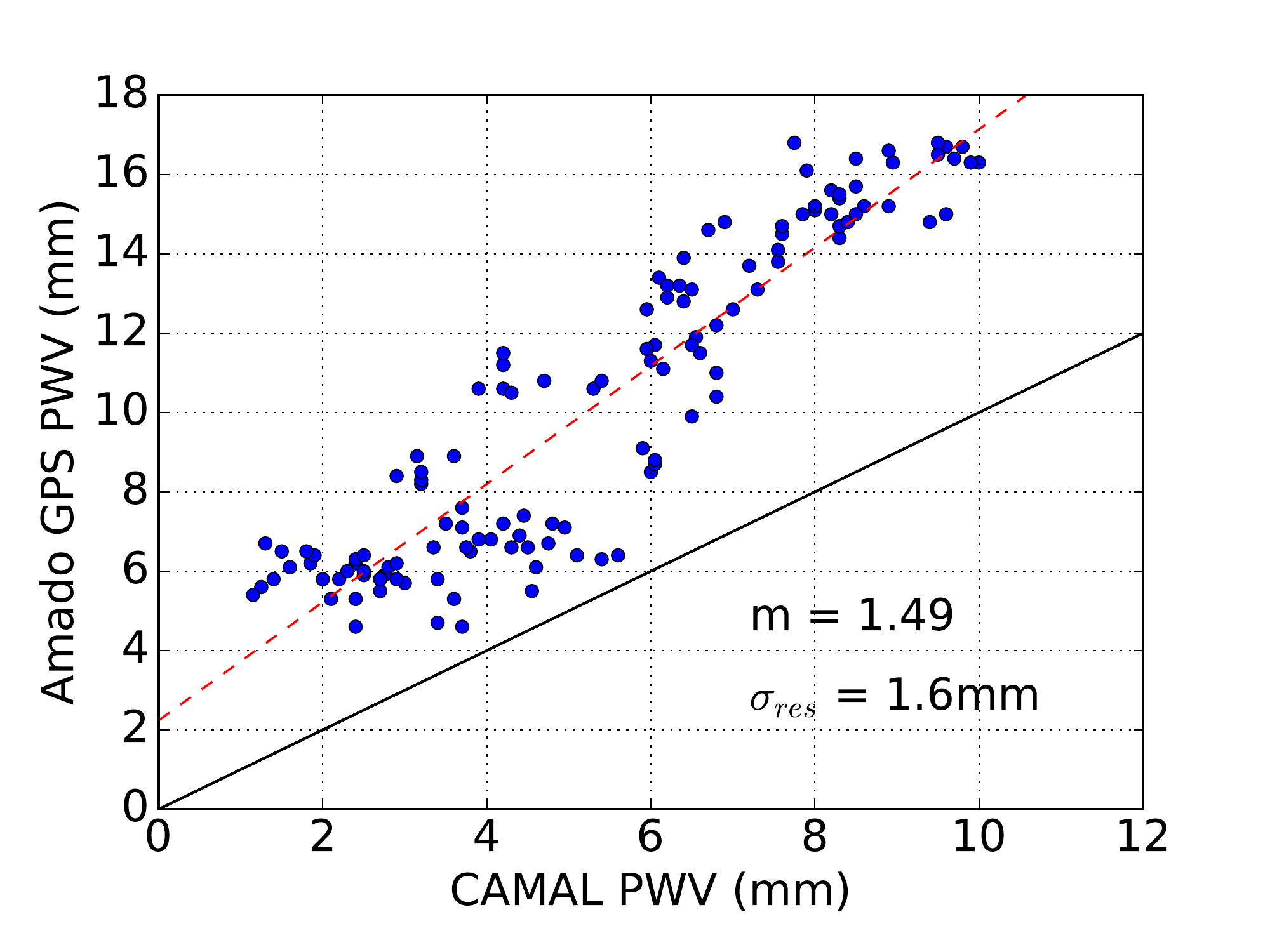}
  \caption{CAMAL PWVs plotted versus Amado PWVs. The black solid line shows the line of one-to-one correspondence, while the red dashed line is a fit to the data. The standard deviation of residuals from the fit is 1.6 mm. }
  \label{fig:azamcamal}
\end{minipage}%
\hfill
\begin{minipage}{.48\textwidth}
  \centering
  \includegraphics[width=0.9\linewidth]{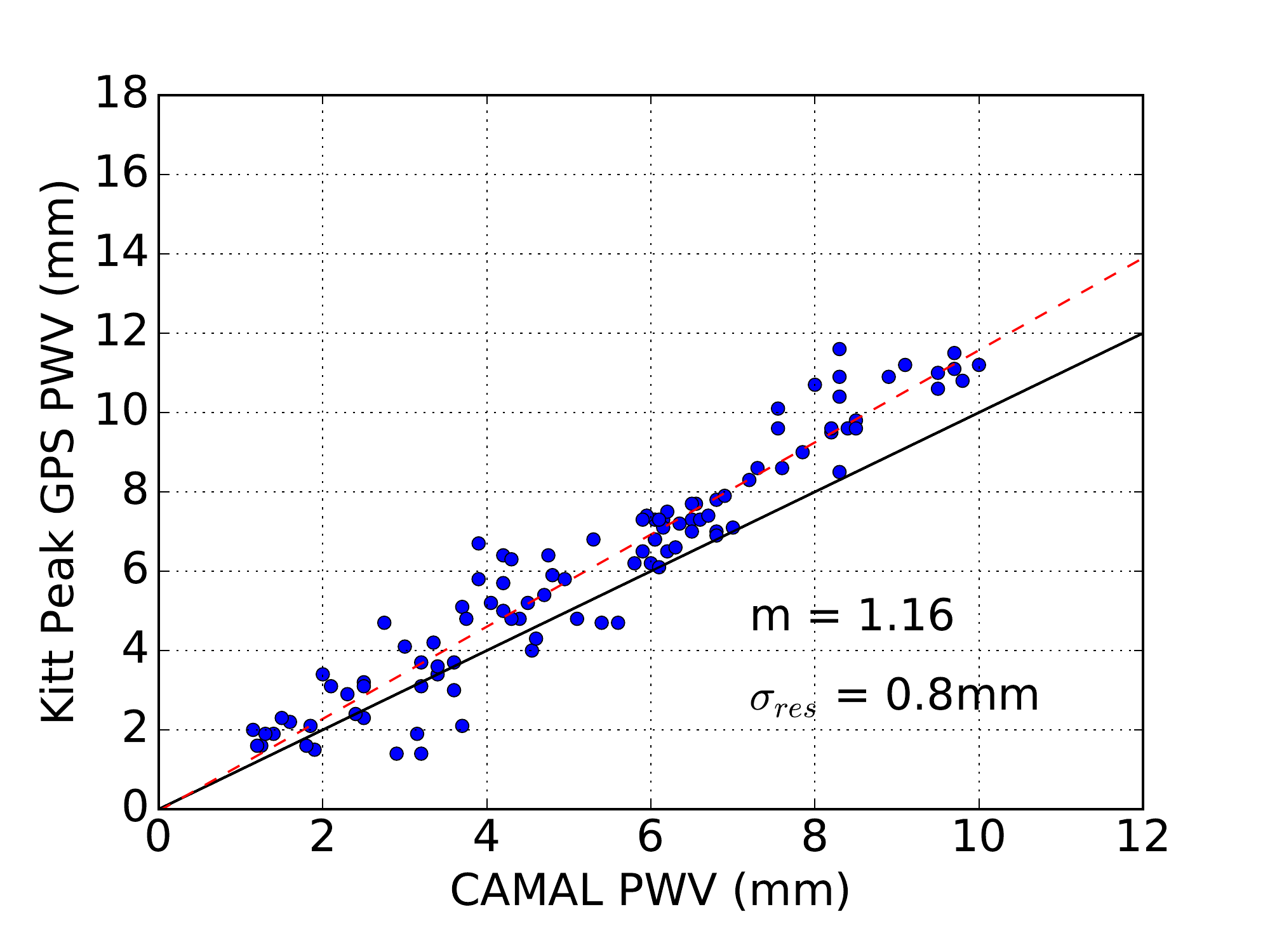}
  \caption{CAMAL PWVs compared to Kitt Peak PWVs. The black solid line shows the line of one-to-one correspondence, while the red dashed line is a fit to the data. The standard deviation of residuals from the fit is 0.8 mm.}
  \label{fig:kittcamal}
  \end{minipage}
\end{figure}

Here we present results using data from the commissioning of CAMAL over four nights in the summer of 2016 and seven full nights in the late fall of 2016. We show in Figure \ref{fig:camaldata} the PWVs extracted following both methods described in \S \ref{sec:extract} for the six nights of data taken in December 2016. In each panel, the gray data points are `real time' PWV measurements from CAMAL colors $c_1$ and $c_2$ using the model grids described in \S \ref{sec:pwvquick}. The gray shaded region is the 3$\sigma$ region from the fit to all the data using a smooth function to describe the time evolution of PWV over the night, as described in \S \ref{sec:fitting}. PWV data from two nearby GPS PWV monitoring systems on the SuomiNet system \citep{Suominet} are also plotted. In red are PWV data from Kitt Peak Observatory and in blue are PWVs from the Amado site.

For the nights shown in Figure \ref{fig:camaldata}, the PWVs found by CAMAL do not vary more than 4 mm in one night. Each night's temporal variation is unique, with some nights showing an increasing trend in PWV, while others show a decrease in PWV over the course of the night. The two different methods for extracting PWV agree well, with no systematic offsets. Times in the night where target stars were changed are noticeable in some nights of data because we typically slewed to a target at high airmass and tracked it upwards towards zenith. The scatter in the photometric data increases after this transition due to the increase in scintillation noise at higher airmass.

\subsection{TRES Comparison}\label{sec:trescompare}
Seven nights of TRES PWV values are compared in Figure \ref{fig:trescamal} to CAMAL values. The gray points are the CAMAL values that were used to determine $A_2$ and the blue points are after $A_2$ was used to calibrate the CAMAL instrument throughput. Because all but two of the TRES and CAMAL measurements are not perfectly contemporaneous (some TRES spectra were taken outside of CAMAL's observing window by as much as a few hours), we plot the CAMAL PWV value nearest in time to the corresponding TRES spectrum taken on the same night. For one night we plot two points because two TRES spectra were taken that night at times separated by a few hours. The values and uncertainties of the CAMAL PWV values are taken from the results of the fit of PWV defined as a spline function, as described in \S \ref{sec:fitting}. These data were used to calibrate the CAMAL instrument throughput at 823 nm. The calibrated model of the instrument throughput results in the excellent correspondence shown here between TRES PWVs and CAMAL PWVs, with a correlation coefficient of 0.98. 

The residuals between CAMAL (post $A_2$ calibration) and TRES values have a root mean square scatter of 0.7 mm. This remaining scatter in the relationship between the CAMAL and TRES data may be due to a combination of spatial and/or temporal variations in PWV between the observations since the corresponding measurements are not all perfectly contemporaneous, and the target stars had a range in angular separation on the sky. The residuals of the CAMAL data relative to the TRES PWV measurements data do not correlate with the temporal or spatial offsets between the two measurements. Our understanding of these residuals will continue to improve with more nights of contemporaneous observations between TRES and CAMAL. Since the TAPAS atmospheric models are directly tied to a physical model of the atmosphere and PWV in an absolute sense, from these comparisons between TRES and CAMAL we conclude that CAMAL produces PWVs with an absolute accuracy of approximately 0.7 mm.

\pagebreak

\subsection{Comparison to Nearby GPS Monitors}\label{sec:GPS}
We investigate how correlated CAMAL PWVs are with the Kitt Peak and Amado PWV monitors by comparing the various sites' PWVs over the nights CAMAL observed. The GPS station at Kitt Peak Observatory is at an altitude of 6900 ft and is approximately 90 km away while the GPS PWV monitoring system in the nearby town of Amado is approximately 30 km away from Mount Hopkins at an altitude of 3000 ft.

In Figures \ref{fig:azamcamal} and \ref{fig:kittcamal} we compare CAMAL PWVs to Amado and Kitt Peak PWVs, respectively.  Error bars are omitted from these plots for clarity, however the GPS water vapor monitors have a reported measurement uncertainty of $\sim$1 mm. In total there were ten nights that had both Kitt Peak and Amado PWV measurements. Since the GPS monitor takes measurements every half hour, we plot all the GPS measurements in a night versus the coincident CAMAL measurements taken as a median over a ten-minute bin of CAMAL PWVs derived from the model grids and centered at the time of the GPS measurement. However, data from two nights in June were excluded in Figure \ref{fig:kittcamal} when the Kitt Peak PWV monitor was reporting extremely high values (\textgreater 25 mm), which are likely due to active precipitation. 

The PWV comparisons between sites show highly correlated conditions between Kitt Peak and Mt. Hopkins with a Pearson correlation coefficient of 0.96 and an RMS difference of 0.8 mm from a linear fit. Amado and CAMAL are slightly less correlated, with a Pearson correlation coefficient of 0.92 and an RMS difference of 1.6 mm from a linear fit. Both Kitt Peak and Amado have wetter conditions when compared to CAMAL values. Due to the elevation differences between the CAMAL site on Mount Hopkins and the GPS sites of Kitt Peak (900 ft lower) and Amado (4800 ft lower), the drier conditions at Mt. Hopkins as shown by CAMAL are expected. The expected average relationship between Kitt Peak and Mt. Hopkins PWVs based on TAPAS atmospheric models, which only differ by site altitude, is that Kitt Peak PWVs are 1.12 times higher than PWVs at Mount Hopkins. This is similar to slope of 1.16 found in the linear fit between CAMAL and Kitt Peak PWVs. 

Correlating PWV between two different sites has been done in previous studies. For example, in \cite{Blake11} the authors were able to correlate two GPS monitors that were 50 km away from each other with a 4000 ft elevation difference. \cite{Blake11} found that the two were highly correlated with deviations from linear at PWV \textless 1 mm. The comparison presented here between Amado and CAMAL also shows deviations from linearity at low PWV values, but this is not the case when comparing Kitt Peak to Mt. Hopkins. The PWV comparisons presented in \cite{Blake11} and this work shows the utility of using one site's PWV measurements to estimate conditions at a nearby site. Selecting PWV data from another site at a similar altitude may be preferable to simply using PWV data from the nearest location.


\section{Conclusions}\label{sec:conc}

We have successfully developed and installed an automatic PWV monitoring system, CAMAL, at the Fred Lawrence Whipple Observatory on Mount Hopkins that will aid the MINERVA, MINERVA-Red, and MEarth exoplanet surveys. CAMAL uses three narrowband filters, one of which is centered on water vapor telluric lines at 823 nm. By observing bright stars throughout a night, changes in flux through the 823 nm on-band filter correspond to changes in telluric water vapor content, or PWV, above Mount Hopkins, while the other two off-band filters serve as a reference to track any achromatic changes in flux not due to water vapor. By constructing a forward model and using independent measurements of PWV derived from high-resolution optical spectra to calibrate CAMAL's throughput, we are able to map the observed CAMAL flux ratios to absolute PWV measurements.

The commissioning data presented in this work spread over 11 nights in the summer and late fall of 2016. With these measurements we show that our photometric precision allows CAMAL to achieve better than 0.5 mm PWV precision for PWV conditions of \textless 4 mm. This level of precision will allow for mmag photometry for transiting exoplanet surveys targeting cool stars in the NIR. Additionally, comparisons between CAMAL PWVs and those from a GPS water vapor monitor at Kitt Peak National Observatory show that the PWVs at these sites are highly correlated with a 0.8 mm RMS difference from a linear fit. This demonstrates the utility of using the PWV conditions from a nearby site at a similar altitude as a first estimate of local PWV.

In future work we plan to upgrade CAMAL to a new design that allows for truly simultaneous measurements of a star through each of CAMAL's three filters. This will involve a total of three optical tubes and CCDs each equipped with one of the three filters that together will simultaneously image the same star. Although the current filter wheel design has allowed CAMAL to achieve its goal of measuring PWVs with $\pm$0.5 mm precision, a simultaneous setup would eliminate CAMAL's current vulnerability to highly variable clouds effectively reducing photometric precision. Additionally, we must currently limit CAMAL to stars brighter than $i=2.5$ mag in order to keep filter cycles periods under 30s to minimize the impact of atmospheric variability due to clouds. With a simultaneous setup, we would not have to limit exposure times and thus could observe fainter stars and increase the number of CAMAL targets. Having more target options could become useful if it is necessary for CAMAL targets to be nearby on the sky with respect to the science targets that will be aided by CAMAL's PWV measurements. Though \cite{Li14} did not find measurable variation on the sky in an 8 day campaign at CTIO, we will perform a similar study to assess the spatial variation of PWV on the sky at Mount Hopkins using CAMAL.

\acknowledgments     
We would like to thank the anonymous referee for his or her helpful comments that improved this manuscript. We also thank the Fred Whipple Observatory support staff for their help machining the base attachment for CAMAL and for their quick on-site support when it was needed. We thank the MINERVA Collaboration for letting us install CAMAL in one of their domes and Jason Eastman for his coding help. We thank David Latham and Allyson Bieryla for picking out and sending us the TRES spectra used in this work. We acknowledge the use of \texttt{astropy} in this work and thank the Astropy Collaboration \citep{Astropy13} for their hours spent working on this open source code. This material is based upon work supported by the National Science Foundation Graduate Research Fellowship under Grant No. DGE-1321851 to A.D.B.


\bibliography{report}   
\bibliographystyle{aasjournal}   
\end{document}